\documentstyle[epsfig]{aipproc}

\begin{document}
\title{Nuclear Equation of State and \\ Internal Structure of Magnetars }

\author{In-Saeng Suh and  G. J. Mathews}
\address{Department of Physics,
        University of Notre Dame, 
        Notre Dame, IN 46556, USA}

\newcommand{\gsim}{\mathrel{\hbox{\rlap{\lower.55ex \hbox {$\sim$}}
                   \kern-.3em \raise.4ex \hbox{$>$}}}}
\newcommand{\be}{\begin{eqnarray}}
\newcommand{\ee}{\end{eqnarray}}
\newcommand{\gcm}{g \, cm^{-3}}
\newcommand{\prl}{Phys. Rev. Lett.}
\newcommand{\prd}{Phys. Rev. D}
\newcommand{\apj}{Astrophys. J.}    

\maketitle

\begin{abstract}
Recently, neutron stars with very strong surface magnetic fields have been
suggested as the site for the origin of observed soft gamma repeaters (SGRs).
We investigate the influence of a strong magnetic field on the properties
and internal structure of such strongly magnetized neutron stars (magnetars).
The presence of a sufficiently strong magnetic field changes the ratio of protons 
to neutrons as well as the neutron appearance density. 
We also study the pion production and pion condensation in a strong magnetic field. 
We discuss the pion condensation in the interior of magnetars as a possible source of 
SGRs.

\end{abstract}

\section*{Introduction}

Recently, observations of the soft gamma repeaters, SGR 0526-66, SGR 1806-20,
SGR 1900+14, SGR 1627-41 and SGR 1801-23 (see \cite{hurley})
with BATSE, RXTE, ASCA, and BeppoSAX have confirmed the fact that these SGRs are
a new class of $\gamma$-ray transients corresponding to strongly magnetized neutron 
stars (magnetars).
Magnetars \cite{duncan,thompson95} are newly born neutron stars
with a surface magnetic field of $B \sim 10^{14} - 10^{15}$ G, probably
created by a supernova explosion.

As relics of stellar interiors, the study of the magnetic fields in and around
degenerate stars should give important information on the role such fields play in
star formation and evolution. However, the origin and evolution of stellar
magnetic fields remains obscure.
The strength of the internal magnetic field in a neutron star in principle could be 
constrained by any observable consequences of a strong magnetic field. For example,
rapid motion of neutron stars may be due to anisotropic neutrino emission
induced by a strong magnetic field \cite{segre}. One could also consider the effect of
magnetic fields on the thermal evolution \cite{heyl} and mass \cite{vshiv} of neutron stars.
Recently, Chakrabarty et al. \cite{chakrabarty} have investigated the gross properties of cold
nuclear matter in a strong magnetic field in the context of a relativistic Hartree model
and have applied their equation of state to obtain masses and radii of magnetic neutron
stars.

Since strong interior magnetic fields modify the nuclear equation of state for degenerate
stars, their mass-radius relation also will be changed relative to that of nonmagnetic stars.
Recently, we have obtained a revised mass-radius relation for magnetic white dwarfs \cite{SM20}. 
For strong internal magnetic fields of $B \sim 4.4 \times (10^{11} - 10^{13}$) G, 
we have found that both the mass and radius increase distinguishably 
and the mass-radius relation of some observed magnetic white dwarfs may
be better fit if strong internal fields are assumed.

If ultrastrong magnetic fields exist in the interior of neutron stars as well,
such fields will primarily affect the behavior of the residual charged particles.   
Standard internal properties such as the nuclear equation of state,
neutron appearance, and the threshold density of muons and pions, would be modified by the 
magnetic field.
Under charge neutrality and chemical equilibrium conditions,
we calculate the ratio of protons to neutrons as well as the pion condensate equation of 
state in the presence of a sufficiently strong magnetic field.
Here we shortly describe and summarize the results.
The details of this work will be published elsewhere \cite{SM99}.

\section*{Inverse $\beta$-decay and neutron appearance in a strong magnetic field}

Let us consider a homogeneous gas of free neutrons, protons, and electrons ($npe$) in
$\beta$-equilibrium \cite{ST} in a uniform magnetic field.
At high densities above $8 \times 10^6$ ${\rm \gcm}$, protons in nuclei are converted
into neutrons via inverse $\beta$-decay: $e^{-} + p \longrightarrow n + \nu$.
Since the neutrinos escape a star, energy is transport away from the system.
Thus, the composition and structure of the star will be modified mainly by inverse
$\beta$-decay. This reaction can proceed whenever the electron acquires
enough energy to balance the mass difference between protons and neutrons,
$Q = m_n - m_p = 1.293$ MeV.
$\beta$-decay is blocked if the density is high enough that
all energetically available electron energy levels in the Fermi sea are occupied.

In order to determine the equilibrium composition and equation of state,
the coupled equations for chemical equilibrium and charge neutrality should now be solved
simultaneously. Figure 1 shows the proton fraction $Y_p = n_p / n_B$, where $n_B$ is the 
baryon density, as a function of the neutron density $\rho_n$ for given field strength,
$\gamma_e = B/B_{c}^{e}$, where $B_{c}^{e} \simeq 4.4 \times 10^{13}$ G.
If the charged particles are in the lowest Landau level, inverse $\beta$ decay is not
suppressed in magnetic fields. 
This means that rapid neutron-star cooling can occur in a strong magnetic field through 
the direct URCA process \cite{leinson}. However, electrons and protons, 
actually, are not in the lowest Landau level for higher densities above a critical density 
from which higher Landau levels begin
to contribute to the chemical potential of electrons and protons, and hence, particle
number densities.
Therefore, discrete Landau levels become continuous and thus the proton concentration
$Y_p$ goes back to the nonmagnetic case as the neutron density increases. As a result,
inverse $\beta$ decay is still suppressed at high densities in strong magnetic fields.

\begin{figure}
\begin{center}
\epsfxsize=7.0cm
\epsfbox{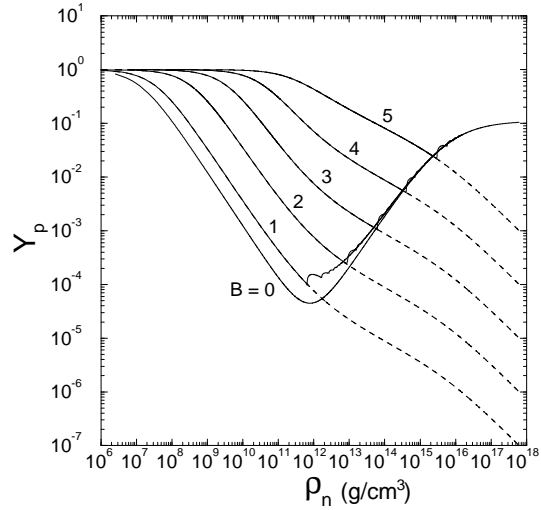}
\vskip 0.2cm
\end{center}
\caption[]{From \cite{SM99}. The proton fraction $Y_p = n_p / n_B$ as a function of the neutron
density $\rho_n$ for given $\gamma_e = B/B_{c}^{e}$'s. 
The $B=0$ line is the non-magnetic case. The dashed lines occur if charged particles
are restricted in the lowest Landau level. 
Numbers 1,2,...,5 correspond to the values of log$\gamma_e$}
\end{figure}

Finally, the equation of state, the mass-energy density 
$\rho = ({\cal E}_e + {\cal E}_p + {\cal E}_n)/c^2$, and
the pressure $P = P_e + P_p + P_n$ (see \cite{lai} for a field strength less than log$\gamma_e$
= 2), are straightforwardly determined.
Figure 2 shows the equation of state for a $npe$ gas in various magnetic fields. 
In this figure we can see that the neutron appearance density for an ideal $npe$ gas increases 
linearly with magnetic field strength.

\begin{figure}
\begin{center}
\epsfxsize=7.0cm
\epsfbox{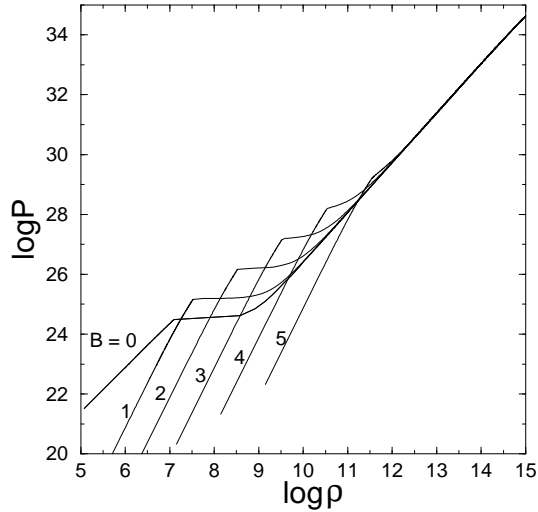}
\vskip 0.2cm
\end{center}
\caption[]{From \cite{SM99}. The equation of state for $npe$ gas in various magnetic fields.
The $B=0$ line correspond to the non-magnetic case. 
Numbers 1,2,...,5 mean the values of log $\gamma_e$.}
\end{figure}

\section*{Pion production and condensation in a strong magnetic field }

At very high density ($\rho \gsim \rho^{\ast}$), neutron-rich nuclear matter
is no longer the true ground state of neutron-star matter. It will quickly
decay by weak interactions into chemically equilibrated neutron star
matter. Fundamental constituents, besides neutrons, may include a fraction of protons,
hyperons, and possibly more massive baryons. 
In particular, if pion condensation exists in a magnetic field \cite{rojas}, 
charged pion production and condensation through $n \rightarrow p + \pi^-$ is possible.

Figure 3 shows the equation of state for an ideal magnetic $npe\pi$ gas with pion condensation. 
We can see that magnetic fields reduce the pion condensation. However, we still have a
distinguishable pion condensate equation of state in strongly magnetized neutron stars.
  
\section*{Discussion}

In this work, we have studied the nuclear equation of state for an ideal $npe$ gas 
in a strong magnetic field. 
Here, we show that the higher Landau levels are significant
at high density in spite of the existence of very strong magnetic fields.
In particular, at high density, the proton concentration approaches the same nonmagnetic 
limit. As a result, the inverse $\beta$ decay is still suppressed in intense magnetic fields.
Therefore, neutron-star rapid cooling is not affected by the direct URCA process
which is enhanced in strong magnetic fields.
Finally, we see that the magnetic field reduces the amount of pion condensation.  
However, we have distinguishable effects of a pion condensate
equation of state in strongly magnetized neutron stars.

It is generally accepted that neutrons and protons in a $npe$ gas are superfluid 
\cite{migdal,yakovlev}.
The charged pion condensate is also superfluid and superconductive \cite{migdal}.
This pion formation and condensation in dense nuclear matter would have the significant
consequence \cite{SM99} that the equation of state would be softened.
First of all, softening the equation of state reduces the maximum mass of the stars.
This softening effect with pion condensation also leads to detectable
predictions \cite{migdal}.
These are: (i) the rate of neutron star cooling via neutrinos would be enhanced,
(ii) a possible phase transition of the neutron star to a superdense state.
(iii) sudden glitches in the pulse period.
In particular, if pion condensation occurs in a strong magnetic field,
it may significantly affect starquakes.

According to the magnetar model by Duncan and Thompson \cite{duncan,thompson95}, 
SGRs are caused by starquakes in the outer solid crust of magnetars. 
In addition, Cheng and Dai \cite{cheng} recently suggested that SGRs may be rapidly rotating
magnetized strange stars with superconducting cores.
Although such models can explain some crucial features, there are still several
unsettled issues \cite{liang}. Therefore, superconducting cores with a charged boson
(pion, kaon) condensate in magnetars might be an alternative model to explain the energy source 
of soft gamma-rays from magnetars.

\begin{figure}
\begin{center}
\epsfxsize=7.0cm
\epsfbox{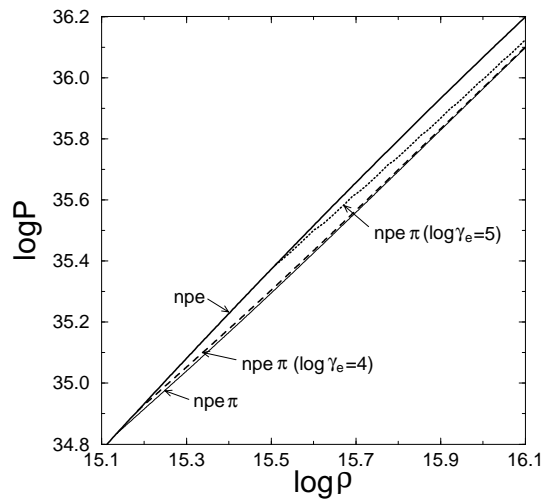}
\end{center}
\caption[]{From \cite{SM99}. The equation of state for an ideal magnetic $npe\pi$ gas
with pion condensation. The thick and thin solid lines are non-magnetic cases (B=0).}
\end{figure}

\section*{ACKNOWLEDGMENTS}
This work supported in part by NSF Grant-97-22086 and DOE Nuclear Theory Grant
DE-FG02-95ER40934.


\begin{references}

\bibitem{hurley} Hurley, K. 2000,
review in this volume (astro-ph/9912061)

\bibitem{duncan} Duncan, R. C. \& Thompson, C. 1992,
\apj ~392, L9

\bibitem{thompson95} Thompson, C. \& Duncan, R. C. 1995,
Mon. Not. R. Astron. Soc. ~275, 255

\bibitem{segre} Kusenko, A. \& Segre, G. 1996,
\prl ~77, 4872

\bibitem{heyl} Heyl, J. S. \& Hernquist, L. 1997,
\apj ~489, L67

\bibitem{vshiv} Vshivtsev, A. S. \& Serebryakoba, D. V. 1994,
Sov. Phys. JETP 79, 17

\bibitem{chakrabarty} Chakrabarty, S, Bandyopadhyay, D., \& Pal, S. 1997,
\prl ~78, 2898

\bibitem{SM20} Suh, I.-S., \& Mathews, G. J. 2000,
\apj ~in press (astro-ph/9906239)

\bibitem{SM99} Suh, I.-S. \& Mathews, G. J.  1999,
Submitted to ApJ. (astro-ph/9912301)

\bibitem{ST} Shapiro, S. L. \& Teukolsky, A. A. 1983,
Black Holes, White Dwarfs, and Neutron Stars (New York: Wiley-Interscience)

\bibitem{leinson} Leinson, L. B., \& Perez, A. 1998,
JHEP ~09, 020

\bibitem{lai} Lai, D. \& Shapiro, S. L. 1991,
\apj ~383, 745

\bibitem{rojas} Rojas, H. Perez 1996,
Phys. Lett. B379, 148

\bibitem{migdal} Migdal, A. B., et al. 1990,
Phys. Rep. 192, 179

\bibitem{yakovlev} Yakovlev, et al 1999, accepted in Physics - Uspekhi, 
(astro-ph/9906456) 

\bibitem{cheng} Cheng, K. S. \& Dai, Z. G. 1998,
\prl ~80, 18

\bibitem{liang} Liang, E. P. 1995,
Astrophys. Space Sci. 231, 69

\end{references}
\end{document}